\documentclass[%
 reprint,
 amsmath,amssymb,
 aps,
]{revtex4-2}

\usepackage{graphicx}
\usepackage{dcolumn}
\usepackage{bm}

\usepackage{subcaption}

\usepackage[table]{xcolor}
\usepackage{booktabs}
\usepackage{enumitem}

\newcommand{\rating}[1]{%
  \begingroup
  \color{blue} 
  \ifnum#1>0 *\fi
  \ifnum#1>1 *\fi
  \ifnum#1>2 *\fi
  \endgroup
}

\usepackage{bbm}

\usepackage{amsthm}
\usepackage{amsmath}

\newtheorem{theorem}{Theorem}

\theoremstyle{definition}
\newtheorem{definition}[theorem]{Definition}

\theoremstyle{remark}

\begin{document}

\preprint{APS/123-QED}
\title{No prior over nuisance parameters is required for credible intervals and Bayes factors} 

\author{Tim Ritmeester, PhD}

\date{\today}


\begin{abstract}
This work shows that there are procedures that maintain the subjective Bayesian property of coherently informing decision making under uncertainty (i.e., if they tell you to take a certain bet, you should) without needing to assign a prior distribution over nuisance parameters, only over the parameter(s) of interest. We first carefully define this sense of 'credibility', and show that within this framework one can condition on any statistic of choice rather than only on the full dataset. This relaxation of credibility is a natural middle ground between Bayesian and frequentist statistics in the following sense: To a $p\%$ Bayesian credible set we should assign a $p\%$ belief after observing both the dataset and the interval, to $p\%$ frequentist sets we can generally only assign a $p\%$ belief \textit{before} observing either the data or the interval, while to the statistical sets proposed here we can assign a $p\%$ belief after observing the set, but not necessarily after inspecting the full dataset ourselves. We then show that choosing a conditioning statistic whose distribution we know, or can bound, for each value of the parameter of interest leads to prior-free Bayes factors even for composite hypotheses, and to non-parametric credible sets with priors only over the parameter of interest. We derive such procedures for several (groups of) statistical problems. Particularly noteworthy is that any frequentist test gives a Bayes factor of the power over the size of the test, and that anytime-validity of E-process-based tests can be carried over to this Bayes factor.
\end{abstract}
\maketitle

\section{Introduction}
Statistical inference ultimately serves decisions under uncertainty, and the two dominant traditions differ in what they offer a decision-maker. A frequentist guarantee concerns a \emph{procedure}: a 95\% interval covers the truth in 95\% of repeated applications \citep{neyman1937}. In decision-theoretic terms, its decision rule is evaluated averaged over independent applications. Its biggest upside is that this evaluation is objective in the sense that everyone should agree to its claims independent of their personal beliefs. It is therefore well suited to scientific or regulatory environments. The downside is that, while making verifiable claims about the performance of the \emph{procedure}, it is not necessarily rational for a decision-maker to follow its recommendations in any particular case.  What one should do in the case at hand depends on base rates and prior beliefs. It is even possible for a procedure to keep its nominal coverage while we can know a specific prediction to be almost certainly wrong (or right) \citep{cox1958,buehler1959}. It is additionally worth mentioning that frequentist procedures are traditionally less flexible and more prone to incorrect usage with regard to optional stopping, multiple comparisons and strength-of-evidence interpretations \cite{lindley1975, mcgrayne2011theory}. Recently developed approaches, particularly those based on E-values, can however significantly close this gap \cite{vovk2021, shafer2021, grunwald2024}. \\~\\
Subjective Bayesian inference addresses the problem of acting rationally in any specific decision. Probability is defined in a subjective, 'belief', sense, through how one would react to any particular uncertain decision. E.g., if you'd accept a bet at $1:1$ odds that a certain event happens, it means you assign at least a $50\%$ probability to that event. Indeed, a set of decision rules is only coherent if it can be derived from a probability ('belief') distribution, and if, upon observing a certain event, they are updated according to Bayes' rule  \citep{definetti1937,teller1973,lewis1999,greaves2006}. In this sense, subjective Bayesian methods really do tell you how to rationally make a decision at hand. If a computation shows that your posterior assigns a probability of $p$ to an event, you should accept any bet for this event at odds better than $(1 - p):p$. 
The main downsides of the subjective Bayesian approach are two-fold: One, the resulting beliefs are subjective, and hence not everyone (those with different priors) will agree that your decisions based on them are wise. Second, it requires you to explicitly declare your prior beliefs. This is notoriously impractical, especially in high-dimensional or non-parametric cases, or in the presence of (many) nuisance parameters \cite{Ghosal_van_der_Vaart_2017}. In practice, heuristic or default priors are often used, doing damage to the 'rational decision-making' interpretation. In one particular situation, both of these downsides are significantly reduced: If we are comparing two simple hypotheses, one may summarize the experiment in a single number independent of the prior, the \textit{Bayes factor} \cite{kass1995}. This signifies the factor by which one should multiply the odds at which one accepts a bet after observing the outcome of this experiment. In practice, however,  there are often nuisance parameters or composite hypotheses requiring implicit or explicit specification of priors, which negates this clean interpretation.\\~\\
Many hybrids between these two traditions have been proposed to try and mitigate their downsides. A first class of such hybrids applies the Bayesian framework, but inserts priors that do not necessarily directly correspond to the user's prior beliefs. The objective Bayes approach uses default priors \citep{jeffreys1946,bernardo1979,berger2006}, empirical Bayes uses a prior derived from data-estimated base rates \citep{robbins1956,efron1973}, and the frequentist-matching approach chooses priors such that credible sets have approximately correct frequentist coverage \citep{welch1963,datta2004}. The underpinning of these methods is further strengthened by the Bernstein-von Mises theorem \cite{vandervaart1998}, implying that in many cases the influence of the prior diminishes to zero with enough data. Although these approaches make the application of Bayesian inference much more practical, in general they lose (for a finite number of samples) the 'rational decision-maker' justification of subjective Bayesian inference. \\~\\
Two approaches that more actively try to maintain the subjective Bayes justification are robust Bayesian analysis \cite{berger1994} and fiducial inference. Robust Bayesian analysis uses \textit{sets} of priors, rather than a single prior, and accordingly reports a set of posteriors. If your prior is in the set of input priors, that means your posterior must also be in the set of output posteriors. This maintains the subjective Bayesian motivation and does not require the user to \textit{exactly} specify their prior. In fact, the approach proposed in the present paper implicitly also uses such a set of priors. Explicit declaration of, and computation with such a set of priors is, however, not in general any easier to bring into practice than the more vanilla subjective Bayesian inference. 
\\~\\
Lastly, fiducial inference \citep{fisher1930, Hacking_Romeijn_2016} is Fisher's attempt to get a subjective belief distribution without specifying a prior. The main idea is that \textit{before} observing any data you should put a $p$-$\%$ belief that any frequentist (say) $p$-$\%$ interval will contain the parameter. Fisher then argues that, if you truly have no prior knowledge about the parameter, then under appropriate symmetry conditions you should think this procedure is equally likely to produce any of the possible intervals. Hence, actually observing the interval produced by the procedure should not change your (prior) belief that such an interval has a $p\%$-probability of containing the parameter, and you should also assign this as your posterior belief. Fiducial inference has since been shown to be inconsistent in many cases \citep{lindley1958}, and in the other cases to correspond to a specific Bayesian prior distribution \cite{Hacking_Romeijn_2016}, decreasing interest in it. It has however, recently seen somewhat of a revival in interest \cite{hannig2016}, and has furthermore been the main inspiration for the present paper. \\

In the present paper, we flip the question around: Can we design procedures, other than the standard Bayesian analysis, that maintain the subjective Bayesian, rational decision-maker, justification? That is, can we design a procedure that returns propositions (e.g. that a parameter falls in an interval, with confidence $p$-$\%$), such that, upon observing such a proposition, we should accept any bet (decision under uncertainty) on this claim at odds better than $(1 - p):p$? In a sense, this achieves a natural middle ground between frequentist and standard subjective Bayesian analysis: \textit{before} observing any particular frequentist claim one should belief that it is/\textit{will be} be correct with the nominal probability, say $p$, but after actually observing the claim one's belief that it is correct may deviate from $p$. To a standard Bayesian claim (at the $p$-$\%$ level) we should assign a $p\%$ belief also after observing both the dataset and the claim, while to the statistical claims proposed here we can assign a $p\%$ belief after observing the claim itself, but not necessarily after inspecting the full dataset ourselves.\\~\\
The structure and main results of this paper are as follows: In Sec.~\ref{sec: credible_propositions} we formalize the above definition of a 'credible' proposition. In Sec.~\ref{sec: construction} we show that indeed such procedures are possible: We are allowed to condition on any statistic rather than on the full dataset. We show that this means that if we can construct an appropriate statistic, we only need a prior over the parameters of interest, rather than also over the nuisance parameters. In Sec.\ref{sec: examples} we give concrete groups of examples where we can make such constructions. For example, we show that we can make non-parametric intervals with one-dimensional priors, that we can construct prior-free Bayes factors for composite and even non-parametric hypotheses, and that rejection of \textit{any} frequentist test results in a Bayes factor. We also show that we can use E-values to allow for anytime-validity of these Bayes factors. Finally, before the final conclusions and discussion, in Sec.~\ref{sec: numerical_verification} we give independent numerical confirmation of the main analytical results made in the previous sections.

\section{Our definition of a credible proposition}\label{sec: credible_propositions}
In the present paper, we address the question: What procedures maintain the subjective Bayesian, rational decision-maker, justification? That is, what procedures can we design that return propositions (e.g. that a parameter falls in an interval, with confidence $p$-$\%$), such that, upon observing this proposition, we should accept those bets (decisions under uncertainty) that the proposition tells us to? We will call such propositions \textit{credible propositions}. \\

In a sense, this achieves a natural middle ground between frequentist and standard subjective Bayesian analysis: \textit{before} observing any particular frequentist claim one should belief that it is/\textit{will be} be correct with the nominal probability, say $p$, but after actually observing the claim one's belief that it is correct may deviate from $p$. To a standard Bayesian claim (at the $p$-$\%$ level) we should assign a $p\%$ belief also after observing both the dataset and the claim, while to the statistical claims proposed here we can assign a $p\%$ belief after observing the claim itself, but not necessarily after inspecting the full dataset ourselves.\\

Let us first give two specific examples before considering a more general form.
For a credible set (e.g., interval), it is as follows.
We will consider a procedure $\mathcal{A}$ that takes as input (some relevant part of) the user's prior beliefs and the full dataset from the experiment ($X$), and returns a (possibly randomized set) $S$. In addition, it specifies some credible level $p$. We will say this procedure returns \textit{credible sets} at level $p$ if, upon observing the returned set, the user should put a $p$ percent belief in the parameter of interest falling within this set. \\~\\
Similarly, for Bayes factors we will consider a procedure $\mathcal{A}$ that takes as input (some relevant part of) the user's prior beliefs as well as the dataset $X$ from the experiment, and returns a non-negative real number (the Bayes factor) for specified hypotheses $H_1$ and $H_0$. We will call the Bayes factors returned by the procedure \textit{credible} if, upon observing this Bayes factor, the user should indeed update the ratio of our belief in $H_1$ and $H_0$ by that Bayes factor. \\

Besides statistical sets and Bayes factors, there are many more propositions that can be relevant, such as the posterior mean or the posterior probability density. We arrive at a more general definition for such 'credible' propositions by the following chain of thought:
\begin{enumerate}
    \item The procedure returning a credible set $S$ (at level $p$) or a Bayes factor $f$ is shorthand for the procedure telling you 'you should put a $p$ percent belief in the parameter being in $S$' and 'you should update your belief by a factor $f$' respectively.
    \item These propositions are valid (i.e., credible) if your posterior conditional on hearing these propositions does indeed put a $p$ percent belief in the parameter being in $S$ and does indeed update your belief by a factor $f$ after hearing those respective propositions.
    \item This means that the above propositions should once again be considered shorthand, for the propositions 'Your posterior conditional on hearing this proposition puts a $p$ percent belief in the parameter being in $S$' and 'Your posterior conditional on hearing this proposition updates your belief by a factor $f$'. 'Credibility' of those propositions is then simply equivalent to them being true.
    
\end{enumerate} 
With this chain of thought we arrive at the following general definition:
\\
\begin{definition}
    \textbf{Credible propositions} \textit{We will call a proposition $Q$  a \textbf{credible proposition} if it gives a true claim about the posterior of the user conditional on observing $Q$ (but seeing nothing about the sample $X$ otherwise).}
\end{definition}

Concretely, this gives the following types of propositions:
\begin{itemize}
    \item \textbf{Credible sets}
    The definition of credible sets $S$ given above is equivalent to the procedure $\mathcal{A}$ returning true propositions of the following form:
    \begin{align}
        q:  \, b(\theta \in s \, |\, Q = q)  = p 
    \end{align}
    for some set $s$ determined by the algorithm.
    \item \textbf{Bayes factors}
    The definition of credible Bayes factors given above is equivalent to the procedure $\mathcal{A}$ returning true propositions of the following form:
\begin{align}
    q: \quad  \frac{b(H_1 | Q = q)}{b(H_0 | Q = q)} = f  \cdot \frac{b(H_1)}{b(H_0)} \,,
\end{align}
for some Bayes factor $f$ determined by the procedure.
     \item \textbf{Posterior probability density} 
    \begin{align}
    q: \quad  b(\theta  \, |\,Q = q ) \, \,  = d(\theta) \quad  \forall \theta 
    \end{align} 
for a density function $d(\theta)$ determined by the algorithm. 

    \item \textbf{Posterior mean}
    \begin{align}
        q: \quad  \int \theta \cdot  b(\theta | Q = q)
     \mathrm{d} \theta = \mu
     \end{align}

\end{itemize}
We will consider more example propositions in what follows, as well as inequality versions of the above propositions.\\

One may expect that any conclusion of standard Bayesian inference, i.e. a proposition about $b(\theta | X)$, to also automatically translate to credible propositions by replacing $b(\theta \in s | X)$ by $b(\theta \in s | Q = q)$. In Sec.~\ref{sec: construction} we will show that for many types of propositions this is true, but not for all of them. \\~\\
Additionally, we observe that for such procedures to be useful we would like them to 'perform well' in the sense that the posterior conditional on $Q = q$, for values of $\theta$ relevant to the application, is close to the belief that we would have assigned if we had inspected the data ourselves, $b(\theta | X)$. 
We will further investigate this subject in Sec.~\ref{sec: numerical_verification}

\section{Conditioning on a statistic instead of the full dataset, and construction with prior only over parameters of interest} \label{sec: construction}

Having defined our weakened notion of credibility (Sec.~\ref{sec: credible_propositions}), in this section we show that we can construct credible propositions such as credible sets and Bayes factors by conditioning on any  statistic of our choosing rather than on the full dataset. In turn, this means that if we have access to (inequalities on) the distribution of this conditioning statistic for each value of the parameter of interest, we can construct such credible propositions about the posterior function using only a prior over the parameter(s) of interest, not over the nuisance parameters.

The approach is to give the procedure access to the data $X$ only through a (possibly randomized) conditioning statistic $T$, rather than the full dataset. Then, the procedure generates some true proposition about the posterior of the user had they observed (only) $T$, replaces the conditioning on $T = t$ by $Q = q$, and returns this proposition as a credible proposition. 

We will define the procedure of conditioning on a statistic as follows: \\
\\
\textbf{Procedure 1: Conditioning on a statistic}
We consider the following procedure $\mathcal{A}$ to construct credible propositions:
\begin{enumerate}
    \item A (possibly randomized) statistic $T$ is computed from the data.
    \item A procedure $\mathcal{A}'$ returns a true proposition $Q'$ about the posterior $b(
\theta | T =t)$ conditional on observing $T$. For example, for a credible interval $s$ this would of the form $Q': b(\theta \in s | T = t) = p$.
    \item The procedure $\mathcal{A}$ then returns the proposition $Q$ obtained by replacing the conditioning on $T = t$ by conditioning on $Q=q$.
\end{enumerate}

The key benefit of this construction over 'standard' Bayesian analysis is that the posterior for the parameter of interest conditional on the conditioning statistic $b(\theta | T = t) \propto b(T = t | \theta)b(\theta)$ depends only on the prior over the parameter of interest and on the likelihood $b(T = t | \theta)$. We are free to choose any conditioning statistic we like, so if we choose a statistic of which we know the distribution for each value of $\theta$ we can construct such credible propositions about the posterior belief using only a prior over the parameter(s) of interest, not over the nuisance parameters. \\

To see that the propositions returned by $\mathcal{A}$ may be credible in the sense of Sec.~\ref{sec: credible_propositions}, the main insight is that given the working of the procedure, $b(T = t | Q = q)$ is only non-zero for values of $t$ for which $q'$, (the proposition $q$ but with conditioning on $Q = q$ replaced by $T =t$) is true. Then:
\begin{align}
    b(\theta | Q = q) &= \int   \, b(\theta | T = t) b(T = t | Q = q) \, \mathrm{d} t \nonumber \\
    &= \int_{t :  q' \text{ is True}} b(\theta | T = t) b(T = t | Q = q)  \mathrm{d} t \nonumber \,.
\end{align}
That is, $b(\theta | Q = q)$ is a weighted sum of $b(\theta | T = t)$ over values of $t$ for which $q'$ is true. From this is it straightforwardly derived, as worked out in more detail in App.~\ref{app: construction}, that Procedure 1 gives true credible propositions for the following cases: 
\begin{itemize}
    \item Posterior density, also when allowing for inequality.
    \item 
     Credible intervals, also when allowing for inequality.
    \item Bayes factors, also when allowing for inequality.
    \item Posterior mean, also when allowing for inequality.
    \item Posterior quantiles (e.g. the median), also when allowing for inequality.
    \item Posterior mean and variance 
    \item Posterior mode.
\end{itemize}
and, interestingly, not in general for the following cases:
\begin{itemize}
    \item Posterior mode inequality
    \item Posterior variance when not also reporting the mean
\end{itemize}
In a trivial sense any credible proposition can be constructed in the manner of Procedure 1, because any credible proposition may itself also be used as a conditioning statistic (in which case this procedure simply returns that proposition). \\~\\

Applying Bernstein-von Mises to the posterior $b(\theta | T = t)$ also allows for investigation of asymptotic 'performance' in the sense of Sec.~\ref{sec: credible_propositions}, that is, whether the resulting posteriors are close to the one we would have had if we would have inspected the dataset ourselves, $b(\theta | X)$. We will use this approach in Sec.~\ref{sec: numerical_verification} to get a feel for this in some concrete examples. \\

Here we can also already conclude that any frequentist hypothesis test, composite/non-parametric or not, directly gives a Bayes factor: Letting $T = 1$ if the test rejects the null hypothesis and $T = 0$ otherwise we quickly get that after observing the outcome of the test $T$ we can report a credible Bayes factor:
\begin{align}
    f &=  \left\{ \begin{array}{ll}
\frac{1 - \beta}{\alpha} & \text{if the test is rejected} \\~\\
\frac{\beta}{1 - \alpha}& \text{otherwise }
\end{array} \right . \label{eq: bf_from_test}
\end{align} where $\alpha$ and $1 - \beta$ are the size of the test and power of the test w.r.t. $H_1$ respectively.\\

\section{Examples} \label{sec: examples}
Applying the methods above to concrete statistical problems requires us to find, for the given problem, a conditioning statistic of which we know the distribution, or have inequalities on it, conditional on each value of the parameter of interest. We identify three groups of statistical problems where we can obtain this, each of which we will discuss in more detail, and give concrete archetypal examples of, below. This allows us to construct non-parametric credible intervals with only one-dimensional priors, and prior-free Bayes factors for composite and non-parametric hypotheses. These groups are as follows:
\begin{itemize}
    \item \textit{Straightforward conditioning statistic with known distribution}\\~\\ Here the conditioning statistic corresponding (almost) directly to the sample estimate of the parameter of interest has a known distribution.
    \item \textit{Frequentist tests converted into Bayes factors} \\~\\ As discussed in Sec.~\ref{sec: construction}, any frequentist test can be converted into a Bayes factor. Of particular interest are composite hypotheses and anytime-valid tests through E-values.
    \item \textit{Pivotal quantities injective in the conditioning statistic, or inequality versions thereof} \\~\\ We show that we may derive conditioning statistics with known distribution if such a pivotal quantity is available, or if we have a similar quantity on which we have pivot-like bounds on the probability density. We show that this can be applied to derive non-parametric credible intervals from probabilistic inequalities.
\end{itemize}
Let us now discuss these cases in more detail.

\subsection{Straightforward conditioning statistic with known distribution}\label{sec: simple_examples}
There are a few cases of parameters of interest and assumptions about the underlying distribution where we can straightforwardly construct a conditioning statistic with known distribution for each value of the parameter.
\subsubsection{Expected fraction of samples satisfying a criterion} \label{sec: cdf_example}
If the parameter of interest $\theta$ is the expected fraction of the samples that satisfy a certain criterion, then one can choose the conditioning statistic $T = \sum_i \mathbbm{1} (\text{$X_i$ satisfies criterion})$ to get $l(\theta | t) \equiv b(T = t | \theta ) = \binom{N}{t} \theta^t (1 - \theta)^{N - t}$, which we can insert in Bayes theorem to obtain $b(\theta | T = t)$ and resulting credible intervals, Bayes factors and other attributes according to Sec.~\ref{sec: construction}. A simple example would be the fraction of a distribution that falls below a certain value $y$, i.e., $\theta = P(X < y) = \mathbb{E}[\mathbbm{1}(X < y)]$.

\subsubsection{Standardized normal mean} \label{sec: pivot_example}
If we know that the samples $X$ are distributed according to a normal distribution with unknown mean $\mu$ and variance $\sigma^2$, but our parameter of interest is only the standardized mean $\theta = \frac{\mu}{\sigma}$, then its rescaled sample estimate $T =\sqrt{N} \frac{\hat{\mu}}{\hat{\sigma}}$ has non-central Student-$t$'s distribution $\sim t_{n-1}(\sqrt{N}\theta)$, from the resulting posterior $b(\theta | T =t)$ conditional on $T$, and resulting credible propositions can be calculated.

\subsubsection{Ignoring information}\label{sec: ignoring_information}
Conditioning on a statistic as in the above examples consists of 'ignoring' part of the information in the sample $X$. In fact, also in standard Bayesian analysis this is in practice done quite regularly. For example, if we'd want to know whether car crashes in a particular area have significantly increased between, say, $2025$ and $2020$, we might like to model the number of car crashes in the area for each year by a Poisson distribution with some prior distribution over its parameter. In doing so, you are conditioning only on the total number of car crashes, and ignoring any other information available about those car crashes: Color and model of the car, number of passengers, etc. You are also ignoring additional information that may not even be logged in any dataset: tire pressure, hours of sleep of the driver, etc. Your full prior might be a complicated joint distribution over all those parameters; actually specifying your beliefs over all known and unknown relevant parameters is impractical, so typically you would just ignore any variables that you don't want to consider, and focus on just the total number of car crashes (or a few other parameters). This is often said to be implicit integrating out of nuisance parameters, and thus implicitly assumes a certain prior on them, not necessarily corresponding to your 'true' prior. \\~\\
The analysis and definition of credibility of this paper thus offer an alternative justification of such conditioning on a subset of information, without assuming anything about the prior over nuisance parameters.

\subsection{Any frequentist test can be used as a bayes factor}
As mentioned in Sec.~\ref{sec: construction}, if we have access to a frequentist test for a null $H_0$ and alternative $H_1$ with size $\alpha$ and power $1 - \beta$, and set $T =1$ when the test is rejected and $T= 0$ otherwise, we quickly get the credible Bayes factor:
\begin{align}
    f &=  \left\{ \begin{array}{ll}
\frac{1 - \beta}{\alpha} & \text{if the test is rejected} \\~\\
\frac{\beta}{1 - \alpha}& \text{otherwise }
\end{array} \right . \label{eq: bf_from_test}
\end{align} 
In practice, if you have access to the exact power and size of the test, you might also have access to the full distribution of the test statistic for that test. If that is the case, one might do better by setting $T$ equal to that, as in the previous example of Sec.~\ref{sec: cdf_example}. On the other hand, if you only have a lower bound $\alpha$ and upper bound $1 - \beta$ of the size and power of the test, respectively, one may use that $\frac{x}{y}$ is increasing and decreasing in $x$ and $y$ respectively (for $x \geq 0$, $y > 0$) and still return $\frac{1- \beta}{\alpha}$ as a credible conservative Bayes factor if the test is rejected. We give two classes of examples where this is useful:

\subsubsection{Bayes factors for composite hypotheses}
If a frequentist test, parametric or not, has a minimum size for any member of the (composite) null-hypothesis, and a minimum power for any member of the (composite) alternative, then one can straightforwardly apply the above to generate Bayes factors \footnote{This can alternatively be phrased as the parameter of interest being $\theta = \mathbbm{1}(\text{$H_0$ is true})$.}. \\

As a straightforward example, consider a casino evaluating whether a roulette player, always betting on red,  is cheating or not. If they were not cheating, their win probability on each spin would be exactly $\frac{18}{37}$ (on European roulette, say at the Monte Carlo). If they were cheating, their win probability is presumed to be $> \frac{1}{2}$. Then the casino may note that the probability that after $N$ spins their win fraction $\hat{p} \equiv \frac{1}{N}\sum_i X_i$ is $p^*$ or higher (where $X_i = 1$ if red wins and $0$ otherwise) is exactly $\alpha \equiv I_{\frac{18}{37}}(p^\ast, N - p^\ast + 1)$ if the player is not cheating, while it is $> 1 - \beta \equiv I_{\frac{1}{2}}(p^\ast, N - p^\ast + 1)$ if they are (with $I$ the incomplete Beta function). Thus, the procedure designed to tell a pit boss whether the player is cheating may return a Bayes factor in favor of the player cheating of $\frac{I_{\frac{1}{2}}(p^\ast, N - p^\ast + 1)}{I_{\frac{18}{37}}(p^\ast, N - p^\ast + 1)}$ if the win fraction is $\geq p^\ast$ (and some other valid lower bound, e.g. $0$, otherwise), and the pit boss should update his beliefs accordingly. We also note that although this specific example is parametric, any non-parametric problem where the parameter of interest is a count (see Sec.~\ref{sec: simple_examples}) will give rise to a Binomial distribution which we can treat in exactly the same way, with the composite hypotheses now encompassing full spaces of distributions. Indeed, in the sence of Sec.~\ref{sec: ignoring_information} we already did this by ignoring any information about the trajectory of the ball, time of betting of the player, etc. 

\subsubsection{Anytime-valid Bayes factors through E-values} \label{sec: e-process}
In the previous example, we used as statistic $T = \mathbbm{1}(\hat{p} \geq p^\ast$) for fixed $p^\ast$, for which we could construct probabilistic bounds through the binomial distribution. Instead, we may use the theory of E-values and E-processes to construct anytime valid or non-parametric versions of such tests. We refer to \cite{vovk2021, shafer2021, grunwald2024} for an introduction to E-values and E-processes. \\~\\
One uses that once one constructs an e-process $E(t)$ for the null-hypothesis one can use Ville's inequality $P(\max_{t} E(t) \geq \frac{1}{\alpha}) \leq \alpha$, while for the alternative we can use that $P(\max_{t} E(t) \geq \frac{1}{\alpha}) \geq P(E(N) \geq \frac{1}{\alpha})$. One can then use the statistic $T = \mathbbm{1}\big(\max_{t \leq N} E(t) \geq \frac{1}{\alpha}\big)$ for pre-specified $\alpha$ to construct the Bayes factor, stopping the experiment whenever it is found out that $\max_{t \leq N} E(t) \geq \frac{1}{\alpha}$. \footnote{Of course, the anytime validity comes from the fact that one may already know this to be the case earlier than $t =N$} \\~\\
In the case of the casino player, we may use an e-process $E(t) = \prod_{i=1}^t \Big(1 + \lambda(X_i - \frac{18}{37})\Big)$, with $- \frac{37}{19} \leq \lambda \leq  \frac{37}{18}$, for the null hypothesis. Note that $P(E(N) \geq \frac{1}{\alpha}) = P\big(\hat{p} \geq \frac{1}{\lambda N} \log(\frac{1}{\alpha}) + (1 - \frac{18}{37}\lambda )\big) \geq I_{\frac{1}{2}}(p^\ast, N - p^\ast + 1)$ under the alternative, where $p^\ast \equiv \frac{1}{\lambda N} \log(\frac{1}{\alpha}) + (1 - \frac{18}{37}\lambda )$. Hence, once it is found out that $T = 1$, one may stop the experiment and alert the pit boss, reporting a Bayes factor of $\frac{1}{\alpha} \cdot I_{\frac{1}{2}}(p^\ast, N - p^\ast + 1)$ in favor of the player cheating. Note that the pit boss should only update their beliefs accordingly if they have no other information (e.g., the number of rounds $t$ at which the experiment was stopped) about the players wins/losses.

\subsection{Pivotal quantities injective in the conditioning statistic, or inequality versions thereof}
Assume that, given some conditioning statistic $T$ and parameter of interest $\theta$, we can construct a quantity $R = r(T, \theta)$ with probability density $f(r)$ independent of $T$ and $\theta$. If additionally $R$ is injective in $T$, then we can use $P(T = t| \theta) = P(R = r(t, \theta)) = f(r(t, \theta)$) and use this to construct credible propositions. \\~\\
As a quick example, if we are keeping track of two Poisson processes with parameters $\lambda_1$ and $\lambda_2$ and we are only interested in their sum $\lambda = \lambda_1 + \lambda_2$, then using as statistic $T$ the waiting time until the $k$-th event, the quantity $2 \lambda T$ is a pivot with Chi-squared distribution, which can be directly inverted. \\~\\ If we only have upper and lower bounds  $f_\pm(r)$ on the probability density that $R(T, \mu) = r$, then one can still generate conservative credible intervals by using:
\begin{align}
    P(\theta \in s | T = t) &= \frac{\int_{s} \mathrm{d}\theta \, b(T = t | \theta) b(\theta)}{\int \mathrm{d}\theta b(T = t | \theta) b(\theta)} \\
    &\geq 
     \frac{\int_{s} \mathrm{d}\theta \, f_+(r(t, \theta)b(\theta)}{\int_{s} \mathrm{d}\theta \, f_+(r(t, \theta)b(\theta) + \int_{\setminus s} \mathrm{d}\theta \, f_-(r(t, \theta)b(\theta)} \nonumber
\end{align}
where in the inequality we used that $\frac{x}{x + y}$ is increasing and decreasing in $x$ and $y$ respectively (for $x \geq 0$, $y > 0$). Here we will give an example of how we use this construction to apply Hoeffding's inequality to generate credible intervals.

\subsubsection{Non-parametric credible interval for mean of bounded variables} \label{sec: mean_example}
Say we have independent samples from a one-dimensional distribution with bounds $\pm1$, and the parameter of interest is the mean $\mu$.
Then we can set $T$ equal to the sample mean plus a random variable with distribution $\text{univ}(-\delta, \delta)$ (a good choice is $\delta \equiv  0.804900 \cdot  \sqrt{N}$ for $p = 0.95$, see App.~\ref{app: derivation_details}), then in App.~\ref{app: derivation_details} we show that we can apply Hoeffding's inequality to construct upper and lower bounds  $f_\pm(r)$ on the probability density that $R(T, \mu) \equiv \mu - T = r$:

\begin{align}
    2 \delta \cdot  f_+(r)  &= \left\{ 
\begin{array}{ll}
 g_-(r) & \qquad \qquad \qquad   \text{if $r \leq - \delta $} \\
g_+(r) & \qquad \qquad \qquad \text{if $r \geq  \delta$} \\
1 & \qquad \qquad \qquad  \text{else}
\end{array}
\right. \label{eq: mean_l_plus}\\
    2 \delta \cdot f_-(r) &= \left\{ 
  \begin{array}{ll}
1 - g_-(r) - g_+(r) & \quad \text{if $ - \delta < r < \delta$} \\
 0 & \quad   \text{else}
\end{array}
\right. \label{eq: mean_l_minus}
\end{align}

where $g_\pm(r) \equiv \exp(-2 (r \mp \delta)^2/N)$.  \\~\\
Thus, upon observing that $T = t$, we get the inequality:
\begin{align}
    P(\mu \in s | T = t) &= \frac{\int_{s} \mathrm{d}\mu \, b(T = t | \mu) b(\mu)}{\int_{-1}^{1} \mathrm{d}\mu b(T = t | \mu) b(\mu)} \\
    &\geq 
     \frac{\int_{s} \mathrm{d}\mu \, f_+(\mu - t)b(\mu)}{\int_{s} \mathrm{d}\mu \, f_+(\mu - t)b(\mu) + \int_{\setminus s} \mathrm{d}\mu \, f_-(\mu - t)b(\mu)} \nonumber
\end{align}
and thus and set $s$ for which this reaches the required credibility can be returned by the procedure. \\~\\
We also note that for the frequentist problem estimating of a mean of bounded variables, tighter confidence intervals than those based on Hoeffding's inequality are available, such as those based on the emperical Bernstein inequality \cite{maurer2009} and recent (anytime-valid!) improvements of confidence intervals based on E-values \cite{waudbysmith2024}. However, conversion of the latter to confidence intervals relies on test-inversion, which we are not allowed to apply here: We have to condition on a single statistic for \textit{all} values of the parameter, not a different one for each (since when forming one's belief one wouldn't 'forget' the conditioning statistics seen for other parameters). In addition, neither directly yields pivot-like quantities injective in a clear conditioning statistic. It would therefore be of interest in future development to find if we can construct pivot-like quantities such as those used here based on those improved inequalities to construct tighter credible intervals.

\section{Numerical verification} \label{sec: numerical_verification}
We have shown analytically that returning propositions that are true conditional on a specified conditioning statistic are credible in the sense of Sec.~\ref{sec: credible_propositions}. In this section we use numerical methods to independently verify this on some of the examples of Sec.~\ref{sec: examples}. We have additionally we have expressed interest in how well these procedures 'perform' (e.g. give tight intervals, close to those we would obtain from standard Bayesian analysis). In this section we also further investigate this for two examples.

\subsection{Validity of credible propositions}
\begin{figure*}[t]
  \centering
  \begin{subfigure}[t]{0.3\textwidth}
    \centering
    \includegraphics[width=\linewidth]{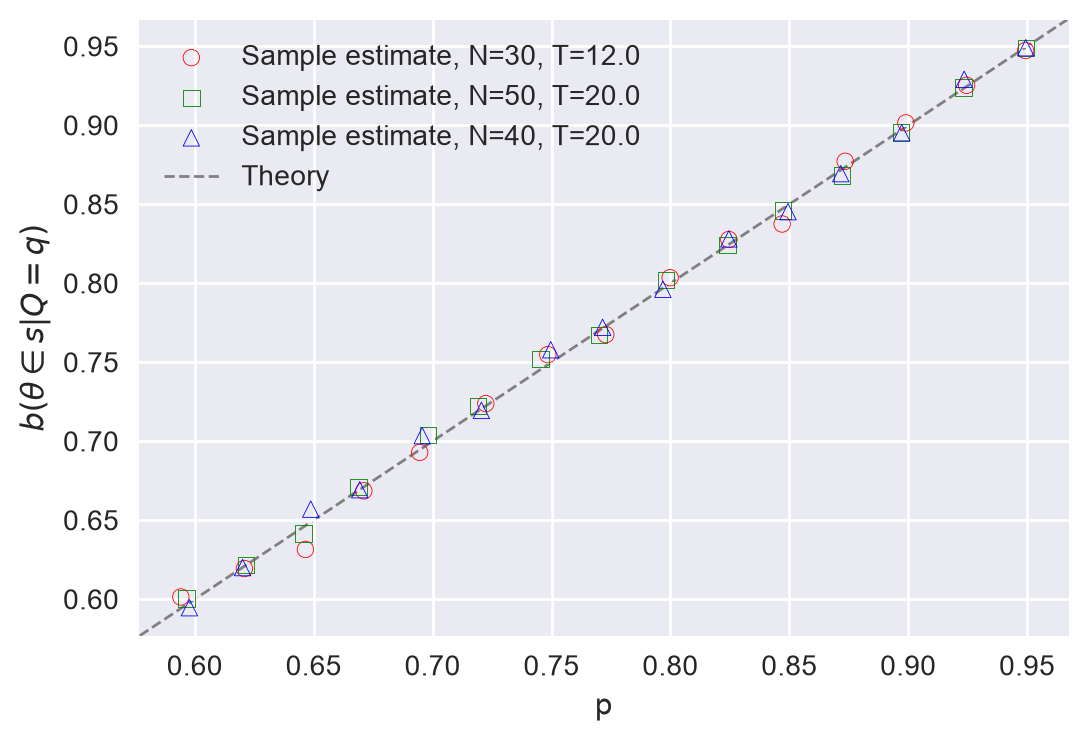}
    \caption{For a credible interval $s$ at various $p$.}\label{fig: CDF_p_vs_b}
\end{subfigure}
\begin{subfigure}[t]{0.3\textwidth}
    \centering
    \includegraphics[width=\linewidth]{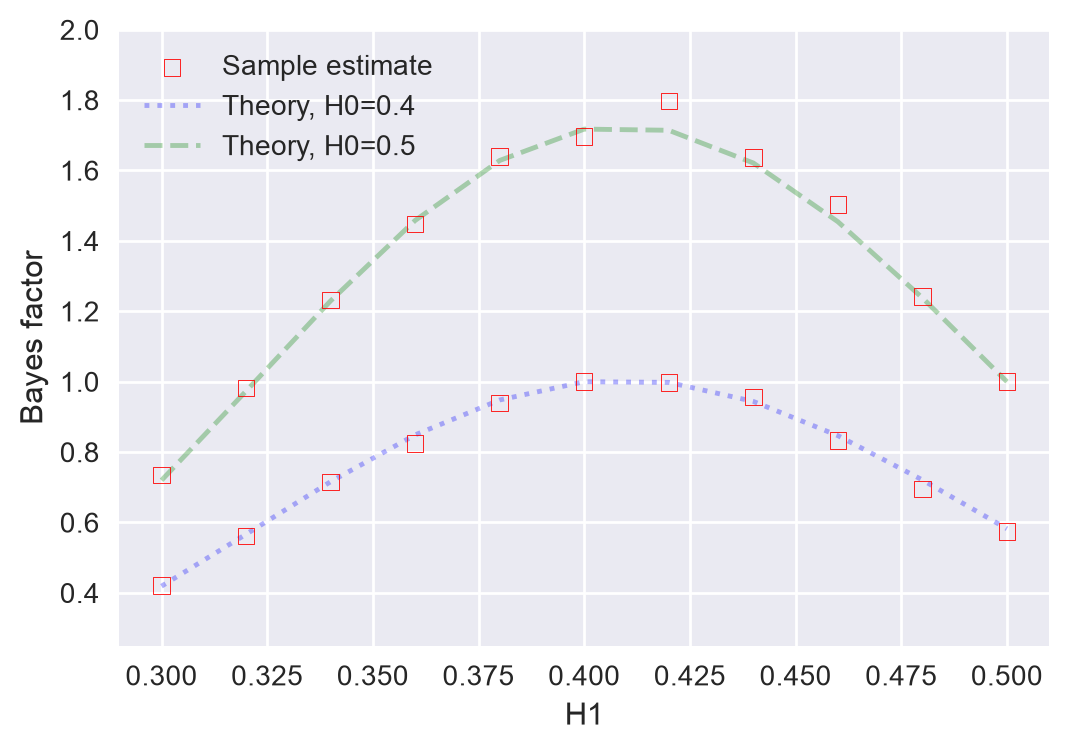}
    \caption{For a Bayes factor between various $H_1$ and $H_0$.}\label{fig: CDF_p_vs_b}
\end{subfigure}
\begin{subfigure}[t]{0.3\textwidth}
    \centering
    \includegraphics[width=\linewidth]{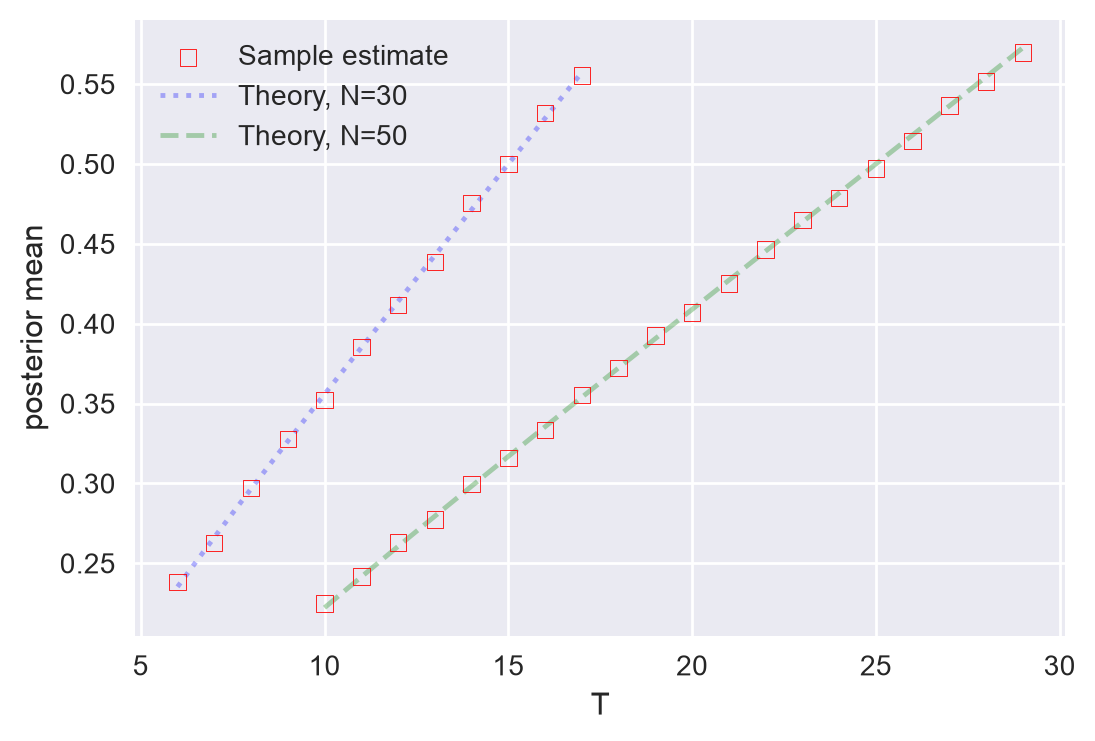}
    \caption{For the posterior mean.}\label{fig: CDF_p_vs_b}
\end{subfigure}
  \caption{Verification of analytical results by an $ABC$ rejection sampler \cite{mikael_2013} for the CDF example of Sec.~\ref{sec: cdf_example}. See Sec.~\ref{sec: numerical_verification} for further details.} \label{fig: cdf_examples}
\end{figure*}

\begin{figure*}[t]
  \centering
  \begin{subfigure}[t]{0.3\textwidth}
    \centering
    \includegraphics[width=\linewidth]{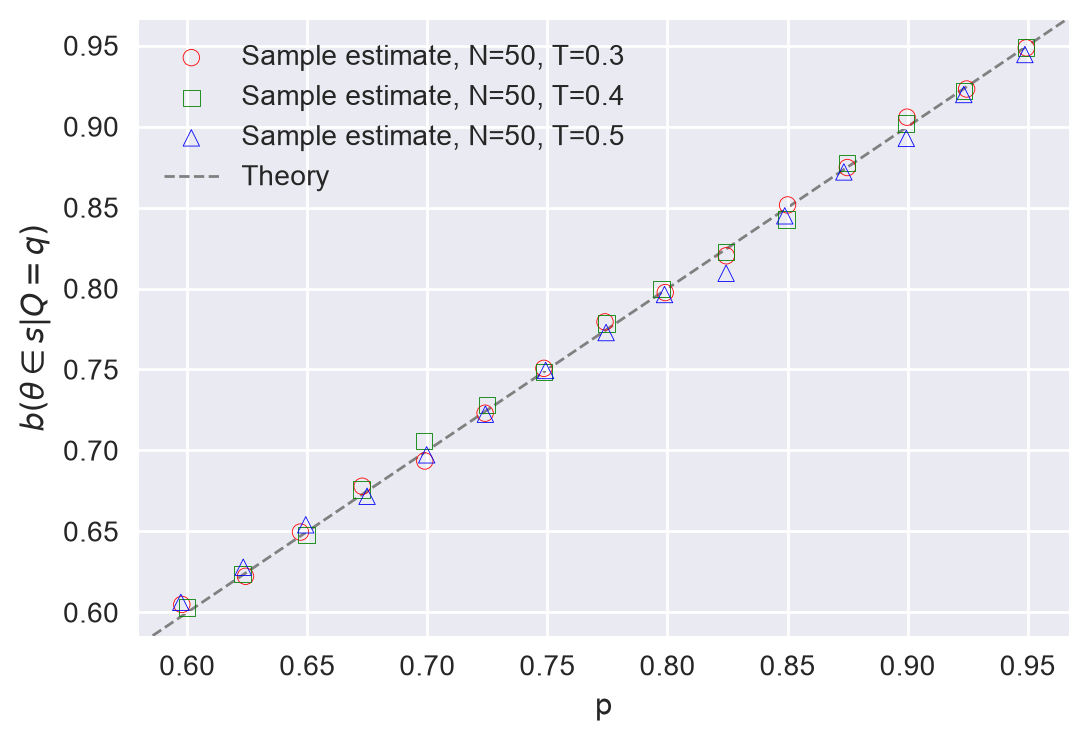}
    \caption{For a credible interval $s$ at various $p$.}\label{fig: CDF_p_vs_b}
\end{subfigure}
\begin{subfigure}[t]{0.3\textwidth}
    \centering
    \includegraphics[width=\linewidth]{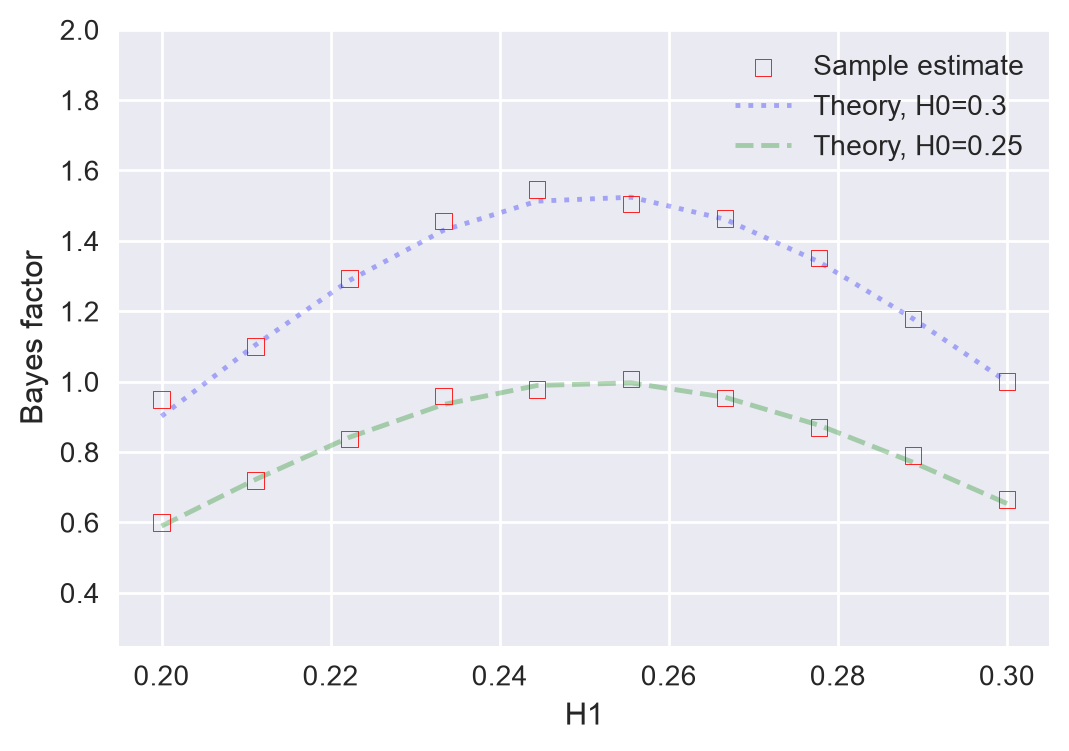}
    \caption{For a Bayes factor between various $H_1$ and $H_0$.}\label{fig: CDF_p_vs_b}
\end{subfigure}
\begin{subfigure}[t]{0.3\textwidth}
    \centering
    \includegraphics[width=\linewidth]{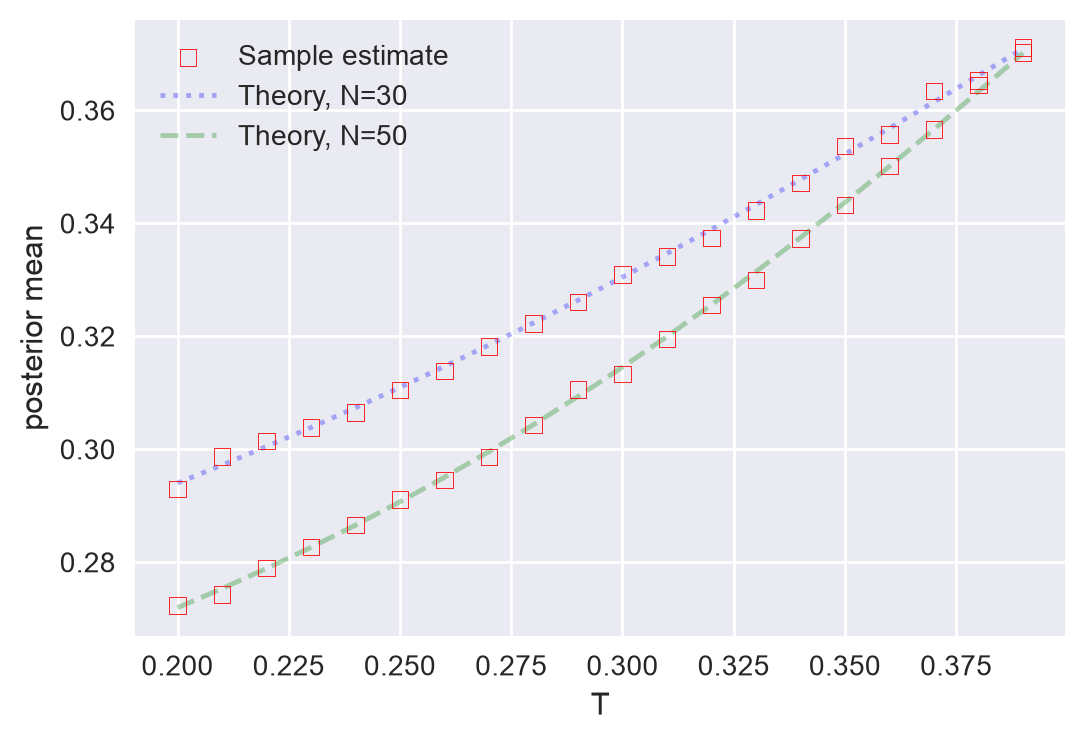}
    \caption{For the posterior mean.}\label{fig: CDF_p_vs_b}
\end{subfigure}
  \caption{Verification of analytical results by an $ABC$ rejection sampler \cite{mikael_2013} for the normal distribution example of Sec.~\ref{sec: pivot_example}. See Sec.~\ref{sec: numerical_verification} for further details.} \label{fig: pivot_examples}
\end{figure*}

\begin{figure}[t]
    \centering
    \includegraphics[width=0.7\linewidth]{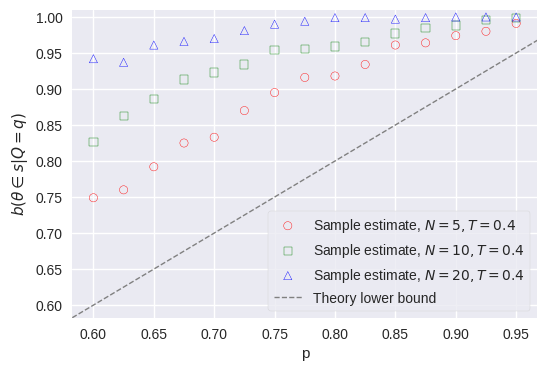}
    \caption{Verification of analytical results by an $ABC$ rejection sampler \cite{mikael_2013} for the mean of bounded variables example of Sec.~\ref{sec: mean_example} for a credible interval $s$ at various $p$. See Sec.~\ref{sec: numerical_verification} for further details.}\label{fig: mean_p_vs_b}
\end{figure}

To obtain independent numerical verification of the validity of credible propositions, we must numerically estimate the posterior conditional on observing a proposition $Q = q$. To this end, we can employ employ an Approximate Bayesian Computation (ABC) rejection sampler. ABC rejection samplers only need access to the (possibly non-parametric) prior distribution over distributions of $X$,  to the procedure $\mathcal{A}$ generating propositions $Q$ from samples $X$, and to the actually observed proposition $q$. By accepting samples from this prior only if $\mathcal{A}$ outputs the observed proposition $Q = q$, and looking at what the value of the parameter of interest was in the cases where this happened, we get samples from the posterior conditional on $Q = q$ and can hence reconstruct it without relying on analytical calculation. For more details on the working of ABC rejection samplers we refer to  \cite{mikael_2013}. Application to credible propositions concerning posterior density is not practically possible because it is such a high-dimensional dimensional object. For lower-dimensional objects such as credible intervals, bayes factors and the posterior mean it is however possible to use an ABC rejection sampler to independently numerically verify our results. We have done so to verify a number of examples, specifically those of Secs.~\ref{sec: cdf_example}, \ref{sec: pivot_example} and \ref{sec: mean_example} in Figs.~\ref{fig: cdf_examples}, \ref{fig: pivot_examples} and \ref{fig: mean_p_vs_b} respectively, for some presumed priors and for different values of $N$ and $T$ (used for obtaining the proposition $q$), $p$ (when concerning intervals) and $H_0/H_1$ (when concerning Bayes factors), showing full agreement with our predictions. For credible intervals we have chosen some different values of $p$, and always chosen intervals symmetric  around posterior mode. We have additionally used a few technical parameters in setting up the sampler, as further discussed in App~\ref{app: numerical_verification}.

\subsection{Performance}
\begin{figure*}[t]
  \centering
  \begin{subfigure}[t]{0.48\textwidth}
    \centering
    \includegraphics[width=\linewidth]{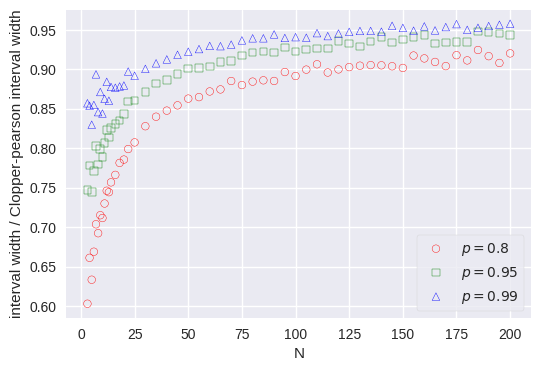}
    \caption{For estimation of the CDF (Sec.~\ref{sec: cdf_example}), with $N = 50$, $m = 0.4 \cdot N$.}\label{fig: CDF_width_vs_N}
  \end{subfigure}
  \hfill
  \begin{subfigure}[t]{0.48\textwidth}
    \centering
    \includegraphics[width=\linewidth]{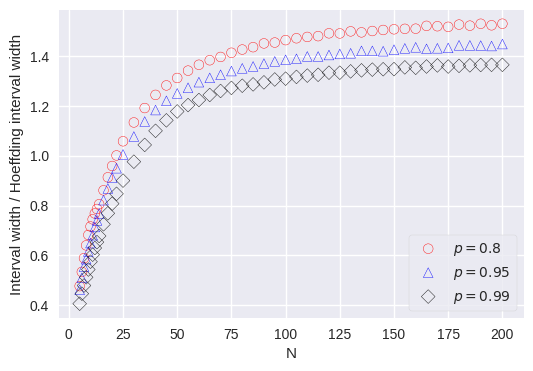}
    \caption{For estimation of the mean (Sec.~\ref{sec: mean_example}), with $N=50$, $m = 0.4$.}
    \label{fig: mean_width_vs_N}
  \end{subfigure}
  \caption{\textbf{Performance}: The width of two proposed credible intervals as a function of the number of samples $N$, compared to the width of standard frequentist intervals for the same estimation (Clopper-Pearson (a) \cite{clopper_1934} and Hoeffding (b) \cite{wasserman2006all}). The frequentist intervals can in turn asymptotically be related to the width of the fully Bayesian estimate; see Sec.~\ref{sec: numerical_verification}. For small $N$ the proposed intervals are narrower, while asymptotically they become (a) equally large or (b) somewhat wider, matching the asymptotic results presented in Sec.~\ref{sec: numerical_verification}.}\label{fig: width_vs_N}
\end{figure*}

We would also like procedures generating credible propositions to perform well in the sense of Sec.~\ref{sec: credible_propositions}, i.e. whether the resulting posteriors are close to the one we would have had if we would have inspected the dataset ourselves, $b(\theta | X)$. Since the standard Bayesian posterior $b(\theta | X)$ uses all available information, performance can be expected to correspond to making the credible sets as small as possible for given $p$ \footnote{Barring peculiar constructions, e.g., ignoring outliers.}. The examples we investigate in more detail numerically, and the behaviour we expect of them, are as follows:

\begin{itemize}

    \item \textit{CDF example:} (Sec.~\ref{sec: cdf_example}) 
    By Bernstein-von Mises the widths of the credible intervals for this example asymptotically approach those of standard frequentist intervals such as the Clopper-Pearson interval, which itself asymptotically matches credible intervals of $b(\theta | X)$, because the conditioning statistic is a sufficient statistic. For finite sample sizes the interval width can be expected to be smaller than frequentist confidence intervals due to them using prior information. For finite sample size the relation to Bayesian intervals depends on the exact prior. This relation of the interval width of these credible intervals compared to the width of Clopper-Pearson intervals is confirmed in Fig.~\ref{fig: CDF_width_vs_N}.

    \item \textit{Mean:} (Sec.~\ref{sec: mean_example}) Since, beyond just conditioning on a statistic, this example also uses probabilistic inequalities we need a few steps beyond Bernstein-von Mises to get their asymptotic width. We show in App.~\ref{app: derivation_details} that asymptotically the interval is, for $p = 0.95$,  $48.79\%$ wider than an equivalent Hoeffding inequality \cite{wasserman2006all} based frequentist interval. How much wider the latter is (asymptotically) than the Bayesian interval depends on the variance of the distribution. In the best case (i.e., maximum variance allowed by the bounded support), it is $38.59\%$ broader than the Bayesian interval, meaning that the present interval would be $2.062$ times broader. For a $99\%$ interval it would be $1.780$ times broader. In Sec.~\ref{sec: mean_example} why at present we didn't use inequalities other than Hoeffding's.\\
    For small samples the intervals produced can once again be expected to be narrower than the frequentist equivalent due to incorporation of prior information. This relation to the width of Hoeffding-based intervals is verified numerically in Fig.~\ref{fig: mean_width_vs_N}.
\end{itemize}

\section{Discussion and conclusion} \label{sec: discussion}
The two dominant traditions of statistics differ in what they offer a decision-maker. A frequentist guarantee concerns properties of a \emph{procedure}. However, it is not always rational for a decision-maker to follow its recommendations in any particular case. Subjective Bayesian inference addresses this by defining probability in a subjective sense, through how one would react to any particular uncertain decision. This comes at the cost of needing to specify a subjective prior belief, which is notoriously impractical when involving high-dimensional nuisance parameters.\\~\\ 
We have shown that there are procedures that maintain the subjective Bayesian property that they can coherently inform decision making under uncertainty (i.e. if they tell you to take a certain bet, you should) but don't correspond to the standard subjective Bayesian inference. We defined this sense of 'credibility' (Sec.~\ref{sec: credible_propositions}), and showed that within this framework one can condition on any statistic of choice rather than only on the full dataset (Sec.~\ref{sec: construction}). We then showed that choosing a conditioning statistic whose distribution we know, or can bound, for each value of the parameter of interest leads to prior-free Bayes factors even for composite hypotheses, and to non-parametric credible sets with priors only over the parameter of interest. We derived such procedures for several (groups of) statistical problems (Sec.~\ref{sec: examples}). Particularly noteworthy is that any frequentist test gives a Bayes factor of the power over the size of the test, and that anytime-validity of E-process-based tests can be carried over to this Bayes factor. We investigated the performance of credible intervals generated by the procedures proposed in this work and found that they are often tighter than equivalent frequentist intervals for small or medium numbers of samples, while either performing equally well or somewhat worse asymptotically.\\

We also observed that the proposed relaxation of credibility is a natural middle ground between Bayesian and frequentist statistics in the sense that: To a $p\%$ Bayesian credible set we should assign a $p\%$ belief after observing both the dataset and the interval; to $p\%$ frequentist sets we can generally only assign a $p\%$ belief \textit{before} observing either the data or the interval; while to the statistical sets proposed here we can assign a $p\%$ belief after observing the set, but not necessarily after inspecting the full dataset ourselves. \\

An important direction for future work is to apply the present method to practical applications, in cases where they can be shown to have benefits compared to the standard constructions. Bayes factors for composite/non-parametric hypotheses seem particularly promising in this regard, since they can be constructed directly from frequentist tests, and eliminate the need for specifying priors completely. \\

Another direction would be to expand and improve upon the list examples derived in this work. The example of a non-parametric mean of bounded variables (Sec.~\ref{sec: mean_example} may certainly be improved upon by applying sharper statistical inequalities or different distributions of noise. \\

Lastly, the philosophical understanding of this work may be deepened through a more extended analysis. This may also have practical benefits. For example, there is a case to be made that only reporting a credible posterior density, and/or the value of the conditioning statistic, should cover any resulting computation of, e.g., credible intervals. In standard subjective Bayesian analysis, propositions about the posterior density are also only interpreted as 'if you would have observed the full dataset, this is what you would believe', and one would not think twice about using an inequality on the posterior mode, even if, by the reasoning of Sec.~\ref{sec: construction}, this inequality does not hold of your posterior after observing that proposition. Similarly, if knowing the conditioning statistic at previous times does not add any information to the conditioning statistic at a later time (as is the case for the counting example of Sec.~\ref{sec: cdf_example}) this may provide a type of anytime-validity different to the of the E-process case (Sec.~\ref{sec: e-process}), \textit{if} it is worked out how the conditioning statistic should be directly incorporated into the credible propositions discussed in this paper.

\section{Acknowledgments}

I would like to thank Prof. Peter Grünwald for his encouragement and insightful recommendations. The suggestions to include an analysis of Bayes Factors and of the incorporation of E-processes are his.

\section{AI Disclosure}
AI was used for code suggestions, literature search, and some more trivial matters such as verification of simple calculations and text suggestions. Other than that AI was not used for this work.

\bibliography{main}       

\begin{appendix}
\section{For which types of credible proposition are we allowed to condition on a statistic?} \label{app: construction}
In Sec.~\ref{sec: construction} we showed that 
\begin{align}
    b(\theta | Q = q) &= \int   \, b(\theta | T = t) b(T = t | Q = q) \, \mathrm{d} t \nonumber \\
    &= \int_{t :  q' \text{ is True}} b(\theta | T = t) b(T = t | Q = q)  \mathrm{d} t \nonumber \,.
\end{align}
That is, $b(\theta | Q = q)$ is a weighted sum of $b(\theta | T = t)$ over values of $t$ for which $q'$ (the proposition $q$ but with conditioning on $Q = q$ replaced by conditioning on $T = t$) is true. We claimed that this allows Procedure 1 to proceduce credible propositions of a number of different forms, which we derive in more detail here. 
\subsection{Posterior density, also when allowing for inequality.}
Here the claim is that $q: \, p^-(\theta) \leq b(\theta  | Q = q) \leq p^+(\theta) \, \, \forall \theta $ is true if
$q': \, p^-(\theta) \leq b(\theta  | T = t) \leq p^+(\theta) \, \, \forall \theta $ is true. This is derived as follows:
 \begin{align}
    b(\theta | Q = q) & \leq\int_{t :  q' \text{ is True}} d^+(\theta)  b(T = t | Q = q)  \, \mathrm{d} t \nonumber \\
    & = d^+(\theta) \int_{t :  q' \text{ is True}}  b(T = t | Q = q)  \, \mathrm{d} t \nonumber\\
    & = d^+(\theta) \int  b(T = t | Q = q)  \, \mathrm{d} t \nonumber\\
    & = d^+(\theta)
\end{align}
with the inequality reversed if $d^+(\theta)$ is replaced by $d^-(\theta)$.
\subsection{Credible intervals, also when allowing for inequality.}
Here the claim is that $q: \, p^- \leq b(\theta \in s | Q = q) \leq p^+ $ is true if $q': p^- \leq b(\theta \in s | T = t)$ is true. This is derived as:
\begin{align}
    b(\theta \in s | Q = q) & = \int_{\theta \in s} \mathrm{d}\theta b(\theta | Q = q) \nonumber\\
    &=\int_{t :  q' \text{ is True}} \mathrm{d}t \int_{\theta \in s} \mathrm{d}\theta \, b(\theta | T = t) b(T = t | Q = q)  \nonumber \\
    &\leq p^+ \int_{t :  q' \text{ is True}} \mathrm{d}t b(T = t | Q = q) \nonumber \\
    &= p^+ \int \mathrm{d}t b(T = t | Q = q) \nonumber \\
    &= p^+ \,,
\end{align}
with the inequality reversed for $p^+$ replaced by $p^-$.

\subsection{Bayes factors, also when allowing for inequality}
Here the claim is that $q: \, f^-  \cdot \frac{b(H_1)}{b(H_0)}  \leq  \frac{b(H_1 | Q = q)}{b(H_0 | Q = q)} \leq f^+  \cdot \frac{b(H_1)}{b(H_0)} $ is true if
    $q': \, f^-  \cdot \frac{b(H_1)}{b(H_0)}  \leq  \frac{b(H_1 | T = t)}{b(H_0 | T = t)} \leq f^+ \cdot \frac{b(H_1)}{b(H_0)}$ is true. This is derived as:
\begin{align}
    b(\theta \in H_1 | Q = q) & = \int_{\theta \in H_1} \mathrm{d}\theta b(\theta | Q = q) \nonumber\\
    &=\int_{t :  q' \text{ is True}} \mathrm{d}t \int_{\theta \in H_1} \mathrm{d}\theta \, b(\theta | T = t) b(T = t | Q = q)  \nonumber \\
    &=\int_{t :  q' \text{ is True}} \mathrm{d}t  b(\theta  \in H_1| T = t) b(T = t | Q = q)  \nonumber \\
    &\leq f^+ \frac{b(\theta \in H_1)}{b(\theta \in H_0)}\int_{t :  q' \text{ is True}} \mathrm{d}t \nonumber \\
    & \quad b(\theta \in H_0 | T = t)b(T = t | Q = q) \nonumber \\
    &= f^+ \frac{b(\theta \in H_1)}{b(\theta \in H_0)} b(\theta \in H_0 | Q = q)\,,
\end{align}
with the inequality reversed for $f^+$ replaced by $f^-$.
\subsection{Posterior mean, also when allowing for inequality}
Here the claim is that $q: \, m^- \leq \int \theta \cdot b(\theta | Q = q) \mathrm{d}\theta \leq m^+ $ is true if $q': \, m^- \leq \int \theta \cdot b(\theta | T = t) \mathrm{d}\theta \leq m^+$ is true. This is derived as:
\begin{align}
    \int \theta \cdot b(\theta | Q = q) \mathrm{d}\theta 
    &=\int_{t :  q' \text{ is True}} \mathrm{d}t \int_{\theta} \mathrm{d}\theta \, \theta \cdot b(\theta | T = t) b(T = t | Q = q)  \nonumber \\
    &\leq m^+ \int_{t :  q' \text{ is True}} \mathrm{d}t b(T = t | Q = q) \nonumber \\
    &= m^+ \int \mathrm{d}t b(T = t | Q = q) \nonumber \\
    &= m^+ \,,
\end{align}
with the inequality reversed for $m^+$ replaced by $m^-$.
\subsection{Posterior quantiles (e.g. the median), also when allowing for inequality}
Here the claim is, for potential bounds on $p$-quantiles $r^\pm$, that $q: \, \int_{-\infty}^{r^-} b(\theta | Q = q) \mathrm{d}\theta \leq p \leq \int_{-\infty}^{r^+} b(\theta | Q = q) \mathrm{d}\theta $ is true if $q': \, \int_{-\infty}^{r^-} b(\theta | T = t) \mathrm{d}\theta \leq p \leq \int_{-\infty}^{r^+} b(\theta | T = t) \mathrm{d}\theta $ is true. This is derived as:
\begin{align}
    \int_{-\infty}^{r^-} b(\theta | Q = q) \mathrm{d}\theta
    &=\int_{t :  q' \text{ is True}} \mathrm{d}t \int_{-\infty}^{r^+} \mathrm{d}\theta \,  b(\theta | T = t) b(T = t | Q = q)  \nonumber \\
    &\leq p \int_{t :  q' \text{ is True}} \mathrm{d}t b(T = t | Q = q) \nonumber \\
    &= p \int \mathrm{d}t b(T = t | Q = q) \nonumber \\
    &= p \,,
\end{align}
with the inequality reversed for $r^+$ replaced by $r^-$.
\subsection{Posterior mode}
Here the claim is that $q: \, \text{argmax}_\theta \, \, b(\theta | Q = q) = m $ is true if $q': \, \text{argmax}_\theta \, \, b(\theta | T = t) = m$ is true. This is derived as:
\begin{align}
b(m | Q = q) &= \int_{t: q' \text{ is True}} \! \! \! \mathrm{d}t \, b(m | T =t)b(T =t | Q=q) \nonumber \\
&> \int_{t: q' \text{ is True}} \! \! \! \mathrm{d}t \, b(\theta | T =t)b(T =t | Q=q) \quad \forall \theta \neq m \nonumber \\
&= b(\theta  | Q=q) \quad \forall \theta \neq m
\end{align}
If the potential credible proposition would only be of the form $\text{argmax}_\theta \, \, b(\theta | Q = q) \geq m$ (i.e. inequality rather than equality), then we can find priors where the inequality in the second line of the derivation is reversed, and hence these type of propositions are in general not valid credible propositions (if produced via Procedure 1).
\subsection{Posterior mean and variance}
For a posterior variance $\sigma^2$ returned by Procedure 1 the situation is different than for the previous cases. We have that:
\begin{align}
\int \theta ^2 b(\theta | Q = q) \mathrm{d}\theta 
    &= \int_{t:q' \text{ is true}} \!\!\!\mathrm{d}t  \int_\theta \theta ^2 b(\theta | T = t) b(T =t | Q = q) \mathrm{d}\theta \nonumber\\
    &= (\sigma^2 + \mu_T^2)\int_{t:q' \text{ is true}} \!\!\!\mathrm{d}t \,   b(T =t | Q = q) \nonumber \\
    & = \sigma^2 + \mu_T^2 \,,
\end{align}
where $\mu_T$ denotes the mean of $b(\theta | T = t)$ over $\theta$. Therefore the variance of $b(\theta | Q = q)$ equals $\sigma^2 + \mu_T^2 - \mu_Q^2$. Since there are many posteriors $b(\theta | T = t)$ for which different values of $T$ share a variance $\sigma^2$ but not a mean, this does not in general ensure that the variance of $b(\theta | Q = q)$ indeed equals $\sigma^2$, and hence propositions about the variance produced by this procedure are not in general valid credible propositions. The situation changes if the algorithm both returns a mean and a variance, in which case all values of $T$ s.t. $q'$ is true also share this mean and hence $\mu_T = \mu_Q$, and the propositions are once again credible.

\section{Parameters used for numerical verification}\label{app: numerical_verification}
For setting up the ABC sampler described in Sec.~\ref{sec: numerical_verification}, we have set the following parameter values:
\begin{itemize}

    \item \textbf{Prior distribution:} For the CDF example (Fig.~\ref{fig: cdf_examples} we have chosen as prior $X \sim \text{N}(\mu, \sigma^2)$ with priors for $\mu$ and $\sigma$ independent and given by $N(0.5, 0.1^2)$ and $\frac{1}{2}\delta(\sigma - 0.1) + \frac{1}{2}\delta(\sigma - 0.2)$, respectively. For the normal distribution example (Fig.~\ref{fig: pivot_examples} we have done the same but with the distribution of $\mu$ truncated to $[0,1]$. For the mean example (Fig.~\ref{fig: mean_p_vs_b} we have done the same, but to ensure bounded support we truncate $\mu$ to $[0.3, 0.7]$ and $X$ to $[\mu - 0.3, \mu + 0.3]$. 

    \item \textbf{Discretization:} We have discretized $\theta$ to $1000$ discrete values with range $[0,1]$. For the numerical verification of the Bayes factor we furthermore found it necessary to smooth the posterior by averaging its value at any point with its $5$ neighbours on either side.

    \item \textbf{Tolerance:} For the normal distribution example we accepted any sample for which the computed quantity (posterior mean, bayes factor or each end of the interval) came within a distance of $0.01$ of the actual value, while for the mean example this was $0.02$ (for the CDF example this was not necessary since the conditioning statistic, and consequently the computed quantities that are returned, are discrete).

    \item \textbf{Number of accepted samples} The number of accepted samples we averaged over was $100.000$ for the Bayes factors, $10000$, $20000$ for the posterior mean of the normal distribution example and $1000$ for the posterior mean of the CDF example, $10000$ for the intervals of the CDF and pivot examples, and $1000$ for the interval of the mean example.
\end{itemize}

\section{Additional analytical details for non-parametric estimation of the mean of bounded variables.} \label{app: derivation_details}

\subsection{Derivation of pivot-like bounds for the mean}
We will here describe the derivation of pivot-like probability bounds for the $T$ of Sec.~\ref{sec: mean_example}. \\
We will use Hoeffding's inequality \cite{wasserman2006all}:

 \begin{align}
    P(\hat{\mu} \geq x) \leq \exp  \big(-2 N (x - \mu)^2 \big)   & & \text{for $x  \geq \mu $}  \\
    P(\hat{\mu} \leq x) \leq \exp  \big(-2 N (x - \mu )^2 \big)   & & \text{for $x  \leq \mu $}
\end{align}

with $\hat{\mu}$ the sample mean.
Writing out $b(T|\mu)$, we get 
 \begin{align}
     b(T|\mu) &= \int_{-\infty}^{\infty} \mathrm{d} \hat{\mu} \,  b(Z = T - \hat{\mu}) b(\hat{\mu}|\mu) \nonumber \\
     &= \frac{1}{2 \delta}\int_{T -\delta}^{T + \delta} \mathrm{d}\hat{\mu} \, b(\hat{\mu} | \mu) \nonumber \\
     &= \frac{1}{2 \delta} \, \cdot \, b(T - \delta \leq \hat{\mu} \leq T + \delta \, | \, \mu )
 \end{align}

 Where we assumed no atoms at the endpoints so that we are free to exchange $<$ with $\leq$. Applying Hoeffding's inequality we get the following upper bounds, defining $g'_\pm(\mu) \equiv \exp(-2 (t \pm \delta - \mu)^2/N)$:

\begin{align}
    2 \delta \cdot b(T = t | \mu) \leq  \left\{ 
\begin{array}{ll}
b(\hat{\mu} \geq t - \delta  \, | \, \mu) \leq g'_-(\mu) & \text{if $\mu \leq t - \delta $} \\
b(\hat{\mu} \leq t + \delta  \, | \,\mu) \leq g'_+(\mu) &\text{if $\mu \geq t + \delta$} \\
1 & \text{else}
\end{array}
\right.
\end{align}

and the following lower bounds:

\begin{align}
    & 2 \delta \cdot b(T = t | \mu) \nonumber \\
    &=  1 - b(\hat{\mu} < t - \delta \, | \, \mu) - b(\hat{\mu} > t + \delta  \,| \, \mu) \nonumber \\
    &\geq \left\{ 
  \begin{array}{ll}
1 - g'_-(\mu) - g'_+(\mu) & \text{if $ t - \delta < \mu < t + \delta$} \\
 0 & \text{else}
\end{array}
\right.
\end{align}

\subsection{Asymptotic analysis}

We can get the asymptotic behaviour of computed intervals by using that $b(T|\theta)$ asymptotically decays very quickly around its mode, so that if the prior is differentiable, we can treat it as approximately constant around this value (assuming  proper scaling of $\delta$). We assume a symmetric confidence interval $[T - \Delta, T + \Delta]$. Plugging in $b(\mu) \propto 1$, and using:

\begin{align}
    \int_{m - \delta}^{m + \delta}\mathrm{d} \mu\, g_\pm(\mu)  &=  \int_0^{2 \delta} \exp(- 2 N x^2) \mathrm{d}x \nonumber \\
    &= \sqrt{\frac{\pi}{8N}} \text{erf} \big(\delta \sqrt{8N} \big)
\end{align}
\begin{align}
    \int_{-\infty}^{m - \Delta }\mathrm{d} \mu \, g_-(\mu) &= \int_{m +\Delta}^{\infty}\mathrm{d} \mu \, g_+(\mu)  \nonumber \\
    &=\int_{\Delta - \delta}^{\infty} \exp(- 2 N x^2 ) \mathrm{d}x \nonumber \\
    &=  \sqrt{\frac{\pi}{8N}} \Big(1 - \text{erf}\big((\Delta - \delta) \sqrt{2N}\big) \Big)
\end{align}

We get, for $\Delta \geq \delta$ and defining $\tilde{\delta} \equiv \delta \cdot \sqrt{\frac{8 N}{\pi}}$ and $\tilde{\Delta} \equiv \Delta \cdot \sqrt{\frac{8 N}{\pi}}$:

\begin{align}
b(\mu \in s | m) &\geq \frac{\tilde{\delta} -  \mathrm{erf}\big( \sqrt{\pi} \tilde{\delta}  \big)}{\tilde{\delta}-\mathrm{erf}\big(\sqrt{\pi}\tilde{\delta} \big) + \mathrm{erfc} \Big(\frac{1}{2}\sqrt{\pi}(\tilde{\Delta} - \tilde{\delta}) \Big)}
\end{align}

Letting the left-hand side equal $p$ and solving for $\tilde{\Delta}$ gives:

\begin{align}
\tilde{\Delta} = \tilde{\delta} + \frac{2}{\sqrt{\pi}} \operatorname{erfc}^{-1} \Big\{ \big(1 - \frac{1}{p}\big)\cdot \Big[\operatorname{erf}(\sqrt{\pi} \tilde{\delta}) - \tilde{\delta}\Big] \Big\} \, \label{eq: delta_mean_eq}
\end{align}

where $\Delta$ is only finite if the argument of the $\text{erfc}^{-1}$ is larger than $0$. We choose $\delta$ to minimize this. Since with this rescaling the equation is independent of $N$, we get that $\delta \propto N^{-1/2}$ and $\Delta \propto N^{-1/2}$. The equation has a unique minimum for $p$ larger than $\approx 0.48$:
\begin{itemize}

    \item The allowed range of $\delta$ is such that the argument of $\text{erfc}^{-1}$ is positive, and the $\text{erfc}^{-1}$ itself is larger than $0$ (since $\Delta \geq \delta$).
    \item On this domain the derivative of $\text{erfc}^{-1}$ increases monotonically from $- \infty$ to $- \frac{\sqrt{\pi}}{2}$. On this domain the derivative of its argument is positive and increasing from $\approx 0.91 \cdot (1/p - 1)$ to $(1/p - 1)$.
    \item Since $- \big(\frac{2}{\sqrt{\pi}} \big) \cdot \big(\frac{\sqrt{\pi}}{2} \big) \cdot 0.91 \cdot (1/p - 1) < 1$ for all $p \geq 0.48\dots$, then by the chain rule and intermediate value theorem there is a single point where the derivative of $\Delta(\delta)$ equals zero. Since $\Delta(\delta)$ is initially decreasing in $\delta$ this must be a minimum.
\end{itemize}

Eq.~\ref{eq: delta_mean_eq} therefore has a unique minimum in $\delta$ for $p$ larger than $\approx 0.48$, and we will choose $\delta$ to be this minimum. \\

Numerical results for a few different values of $p$ are shown in Tab.~\ref{tab: delta_p_relation}. The equivalent Hoeffding-based frequentist intervals have half-widths of $\sqrt{\log(2/(1-p))}/\sqrt{2N}$, so that we get that asymptotically the intervals that we compute here are $58.24\%$, $48.79 \%$ and $40.82 \%$ larger for $p = 0.8$, $p = 0.95$ and $p=0.99$ respectively.

\begin{table}[h!]

\centering

\begin{tabular}{lcc}
\toprule
$\mathbf{p}$ & $\mathbf{\delta/\sqrt{N}}$ & $\sim \! \mathbf{\Delta/\sqrt{N}}$ \\
\midrule
0.80\, \, & \,0.865069 \,\,& 1.699445 \\
0.95 \,\, & \,0.804900 \,\,& 2.020716 \\
0.99 \,\, & \,0.773710 \,\,& 2.292040 \\
\bottomrule

\end{tabular}

\caption{Asymptotic interval half-width $\Delta$ of the interval for estimation of the mean, and asymptotically optimal $\delta$, for different values of $p$. }

\label{tab: delta_p_relation}

\end{table}

\end{appendix}

\end{document}